\documentclass[a4paper,11pt]{article}

\usepackage{style_file}

\usepackage{upgreek}
\usepackage{url}
\usepackage{adjustbox}
\usepackage[normalem]{ulem}
\usepackage{booktabs} 

\begin{document}

\title{Hanbury Brown-Twiss interference with massively parallel spectral multiplexing for broadband light}

\author[1, \ast]{Sergei Kulkov}\note{Corresponding author. sergei.kulkov@fjfi.cvut.cz}

\author[1]{Ondrej Matousek}
\author[1]{Dmitrij Sevaev}

\author[2]{Stephen~Vintskevich}
\author[2]{Yury Kurochkin}

\author[1]{Lou-Ann Pestana De Sousa}
\author[1]{Lada Radmacherova}

\author[3]{Ermanno~Bernasconi}
\author[3]{Claudio Bruschini}
\author[3]{Tommaso Milanese}
\author[3]{Edoardo Charbon}

\author[1,4]{Peter~Svihra}
\author[1,4,5]{Andrei Nomerotski}

\affiliation[1]{Faculty of Nuclear Sciences and Physical Engineering, Czech Technical University, Břehová 7, Prague, 11519, Czech Republic}

\affiliation[2]{Quantum Research Center, Technology Innovation Institute, Abu Dhabi, 9639, UAE}

\affiliation[3]{AQUA Laboratory, École polytechnique fédérale de Lausanne, Rue de la Maladière 71, Neuchâtel, CH-2000, Switzerland}

\affiliation[4]{Institute of Physics of Czech Academy of Sciences, 
Na Slovance 1999/2, Prague, 18200, Czech Republic}

\affiliation[5]{Department of Electrical and Computer Engineering, Florida International University, 10555 West Flagler st., Miami, 33174, FL, USA}

\abstract{

Two-photon interference is a fundamental resource for quantum technologies and optical quantum computing, underpinning precision measurements, scalable entanglement distribution, and the operation of photonic circuits and quantum network protocols. Here, we report the first demonstration of massively parallel, wavelength-resolved photon bunching, revealing Hanbury Brown-Twiss correlations across 100 independent spectral channels. These observations are enabled by a fast, data-driven single-photon spectrometer that achieves 40 pm spectral and 40 ps temporal resolution over a 10 nm bandwidth, providing simultaneous access to spectro-temporal photon correlations without the need for narrowband filtering. This approach preserves photon flux while enabling high-dimensional quantum interference measurements across a broad spectrum. Our results establish frequency-multiplexed two-photon interference as a scalable and throughput-efficient platform for quantum-enhanced photonic technologies, offering a practical route toward room-temperature architectures that overcome loss limitations and advance the scalability for a variety of applications.
}


\maketitle

\section{Introduction}\label{sec1}

Two-photon correlation analysis stands as a cornerstone of contemporary quantum optics, tracing its origins back to foundational studies by Glauber \cite{glauber1963coherent&incoherentstates, glauber1963quantum_optical_coherence}. This powerful method has found diverse applications across numerous fields, including super-resolution microscopy \cite{Degen2017, Ndagano2022}, quantum communication protocols \cite{Samara2021}, two-photon resonance fluorescence \cite{Zubizarreta_Casalengua2024-ew, Varnavski2023}, stellar intensity interferometry (SII) \cite{HBT_original, dravins2015long, rivet2018optical, Horch2025, guerin2025stellar}, and its advanced phase-sensitive variants \cite{Gottesman2012, Stankus2022, Baryakhtar2024, Tang2025}. Central to these applications are distinct correlation signatures between photons arising from fundamental quantum interference effects, predominantly characterized by the indistinguishability of photons as in the Hanbury Brown-Twiss (HBT) \cite{HBT_paper2, HBT_paper3} and Hong-Ou-Mandel (HOM) \cite{HOM_effect} phenomena. 

Quantum interference is also a fundamental building block of optical quantum computing, underlying gate operations, photonic circuits, and quantum network protocols that connect distributed quantum processors \cite{Kok2007, Slussarenko2019}. However, scalability in these systems is critically limited by optical loss, which constrains both communication distance and computational fidelity. A promising strategy to mitigate these limitations is multiplexing, where multiple parallel attempts --- such as simultaneous Bell-state measurements --- improve entanglement generation rates and overall network reliability. In this context, massively parallel, wavelength-multiplexed two-photon interference provides a practical route toward scalable, room-temperature quantum photonic architectures. By enabling many concurrent correlation measurements, this approach supports measurement-based quantum computing and quantum repeater networks, both of which rely on efficient entanglement swapping.

\begin{figure}[h!]
    \centering
    \includegraphics[width=0.92\linewidth]{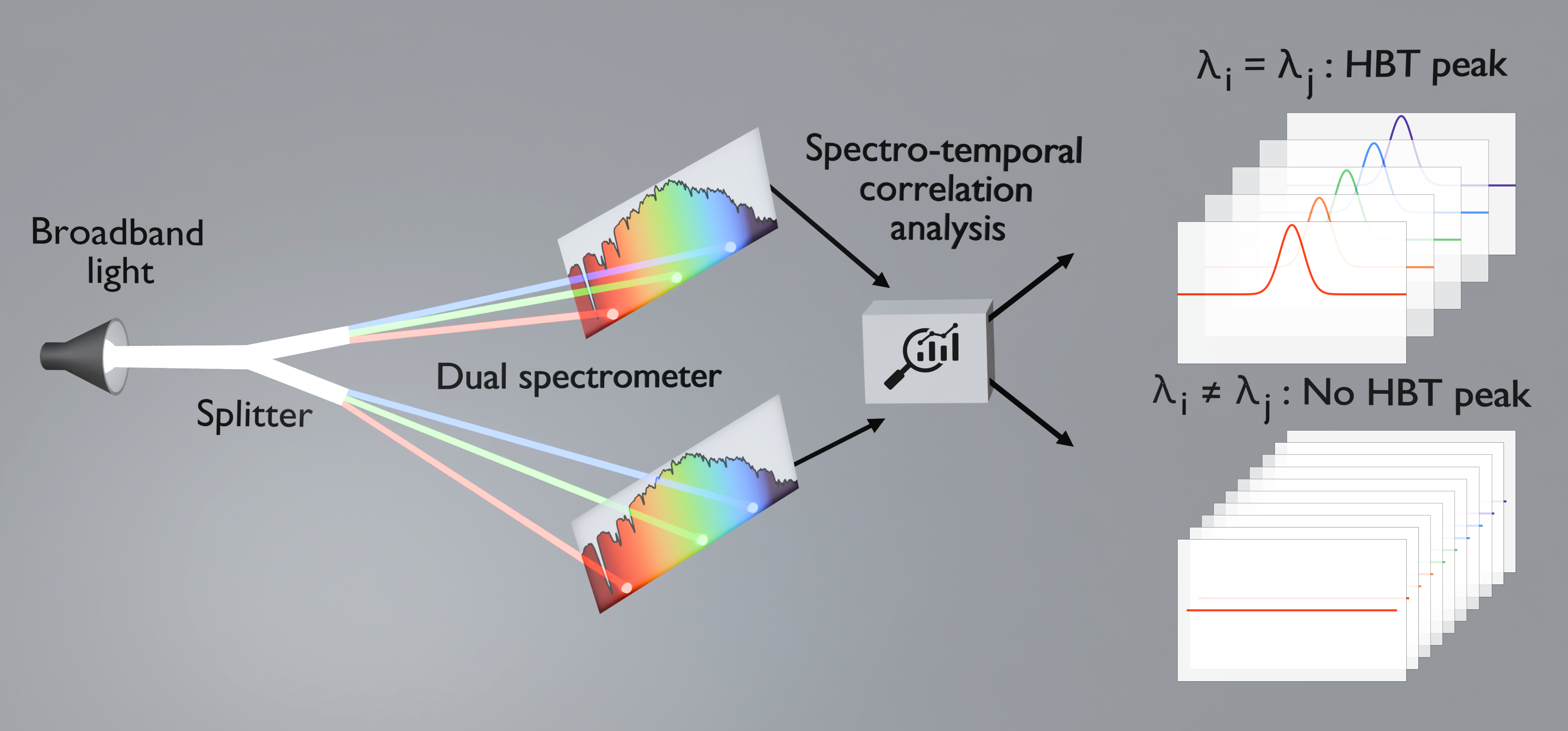}
    \caption{Conceptual illustration of wavelength-resolved two-photon interference. Broadband thermal light is split into two beams and analyzed using a dual-arm spectrometer, producing time- and wavelength-resolved single-photon spectra for each path. A spectro-temporal correlation analysis is performed between spectral channels on both sides. When photons occupy the same spectral bin $(\lambda_i = \lambda_j)$, quantum bunching leads to an HBT peak --- an enhancement in coincidence counts at small time delays between the two photons. In contrast, mismatched spectral bins $(\lambda_i \neq \lambda_j)$ yield flat, uncorrelated distributions, as expected for distinguishable photons. This setup enables massively parallel observation of two-photon interference across a wide spectral range.}
    \label{fig:fig1_idea}
\end{figure}

In this study, we directly measure and quantify the indistinguishability of the photons employing a fast spectrometer combined with the precise analysis of spectral and temporal data performed during post-processing, as schematically presented in Figure \ref{fig:fig1_idea}. For the first time, we unambiguously observe the HBT effect of photon bunching in broadband light simultaneously across 100 spectral bins in a continuous 10~nm range, establishing a new methodology for massively multiplexed correlation sensing over an extensive frequency range, with the potential of substantially improving the resolution, sensitivity, and scalability for quantum optical measurements.

A particularly compelling application of frequency-multiplexed two-photon techniques aims at achieving superior angular (astrometric) resolution in astronomy. Traditionally, interferometric methods have relied on single-photon interferometry, which requires precise photon phase measurements of light emitted by astronomical sources detected by telescope stations positioned along extensive baselines. However, achieving large baselines with single-photon interferometry is technically challenging due to the stringent requirement for phase-stable optical links \cite{tenBrummelaar2005, Pedretti2009}. In contrast, techniques based on the two-photon HBT effect, historically employed in SII to measure stellar diameters, inherently bypass this requirement. Recent proposals suggest further enhancements using quantum-mechanically entangled photon pairs, which could eliminate the need for phase-stable links altogether, thus enabling arbitrarily long baselines \cite{Gottesman2012, harvard1, harvard2, Stankus2022, Brown2023}. This quantum-enhanced approach holds transformative potential for astronomical observations by dramatically improving achievable resolutions. Moreover, frequency multiplexing can amplify these benefits, facilitating simultaneous and independent measurements across multiple spectral bins, subsequently combined to enhance measurement sensitivity significantly. Notably, this multiplexed methodology is equally advantageous for conventional SII techniques.

Another application for the spectrometer with excellent spectral and temporal resolutions is resonance fluorescence, where precise two-photon interference and correlation measurements can yield groundbreaking insights into quantum optical processes in atomic and molecular physics. Spectral correlation patterns observed in resonance fluorescence typically manifest as distinct photon bunching lines and anti-bunching circles, unveiling quantum behaviour at off-resonant frequencies \cite{Mukamel2020-ul, Zubizarreta_Casalengua2024-ew}.

Wavelength-multiplexed two-photon interference is a critical resource for boosting the throughput of quantum communication networks. In widely used entanglement swapping protocols, two spontaneous parametric down-conversion (SPDC) sources each generate a pair of entangled photons. One photon from each pair is directed to a central station where a HOM interference measurement projects entanglement onto the two distant nodes \cite{PhysRevLett.71.4287}. These protocols are foundational for linking quantum processors and for enabling device-independent quantum key distribution (QKD).
However, the probabilistic nature of SPDC imposes a fundamental trade-off between brightness and fidelity: increasing the pair production rate raises the likelihood of double-pair emissions, which degrade entanglement quality. Additionally, SPDC sources typically emit broadband spectra, requiring narrow spectral filtering to ensure photon indistinguishability and high interference visibility \cite{Li2023}. Such filtering, however, severely limits the usable photon flux and network throughput.
To overcome this, we propose treating a broadband SPDC source as a collection of parallel, independent wavelength channels --- each producing low-probability, Fourier-limited single-photon pairs. By leveraging wavelength multiplexing instead of spectral filtering, this approach can substantially increase the number of successful entanglement swaps per unit time. This parallelization strategy represents a promising route toward scalable, high-rate, all-photonic quantum repeaters \cite{Azuma2015}.

Previous attempts to measure the frequency-multiplexed HBT effect were constrained by either limited temporal or spectral resolution or by a narrow frequency range with few spectral bins. Earlier investigations of spectrally resolved HOM interference employed slower cameras, utilizing nanosecond-scale resolution to identify temporal coincidences in photon pairs. Temporal coherence was usually maintained by precisely delaying one photon relative to the other using mechanical stages, effectively enabling sub-picosecond precision and thereby providing good visibility for HOM interference \cite{Zhang2020, Svihra2020, Zhang2021}. More recent efforts, such as those described in \cite{Dilena2025, Silva2016, Ferrantini2025}, achieved significantly improved timing resolution, 20--40 ps. However, these studies still faced limitations, with only a small number of spectral channels, eight channels in \cite{Dilena2025} and five narrow neon spectral lines in \cite{Ferrantini2025}, or restricted spectral coverage, less than 0.3 nm, when investigating two-photon transitions in polariton condensates \cite{Silva2016}. Similarly, SII experiments typically implemented frequency binning through multiple narrow-band filters and standalone photon detectors, thereby limiting scalability \cite{Vogel2024, Leopold2024,tolila2024increasing, Horch2025}.

Other existing techniques, which could achieve simultaneously good spectral and temporal resolutions for the frequency multiplexing, employ dispersive fibers to translate changes in time delay to the photon frequency \cite{Cohen1985}, spectral phase shaping using spectral shear via electro-optic modulation \cite{davis2018experimental}, and an approach based on chirped fiber Bragg gratings \cite{davis2017pulsed}. These systems are interesting in their own right, but typically require complex infrastructure for pulsed operation and synchronization, and do not provide multi-photon detection capabilities as observation of the two-photon interference was not their primary motivation.

Maximizing interference contrast, or equivalently, visibility, necessitates simultaneously achieving exceptional temporal ($\sim$10 ps) and spectral ($\sim$10 pm) resolutions, thereby approaching the fundamental limits set by the Heisenberg Uncertainty Principle \cite{heisenberg1949physical}, also known as the Gabor or Fourier or Heisenberg-Gabor limits \cite{Gabor1946}. Current advances in multi-pixel single-photon detector arrays enable building of wideband spectrometers and also address challenges related to temporal resolution, which commonly limits the achievable contrast due to peak smearing in HBT measurements. These innovative data-driven detectors significantly enhance the practical applicability of quantum optical phenomena across various real-world scenarios
by enabling simultaneous spectro-temporal two-photon correlation measurements without narrowband filtering. The presented approach provides a generally applicable experimental framework that is relevant not only for the mentioned quantum networks and stellar intensity interferometry, but also for a range of related applications, including thermal-light range sensing, pseudo-natural-light (anti)superbunching and ghost-imaging protocols, as well as studies of superthermal photon statistics in bimodal nanolasers \cite{PhysRevA.95.053809, PhysRevX.8.011013, Ye:22, PhysRevApplied.20.014060}.

In the following, we describe the results demonstrating the HBT effect of two-photon interference for broadband light, post-selecting photons with the same frequency and timestamps as determined by precise measurement in the spectrometer. We then discuss the relevance of the results to several scenarios of two-photon interference with projections based on the achieved performance.

\section{Materials and methods}

\begin{figure}[h]
\centering
\includegraphics[width=1.0\textwidth]{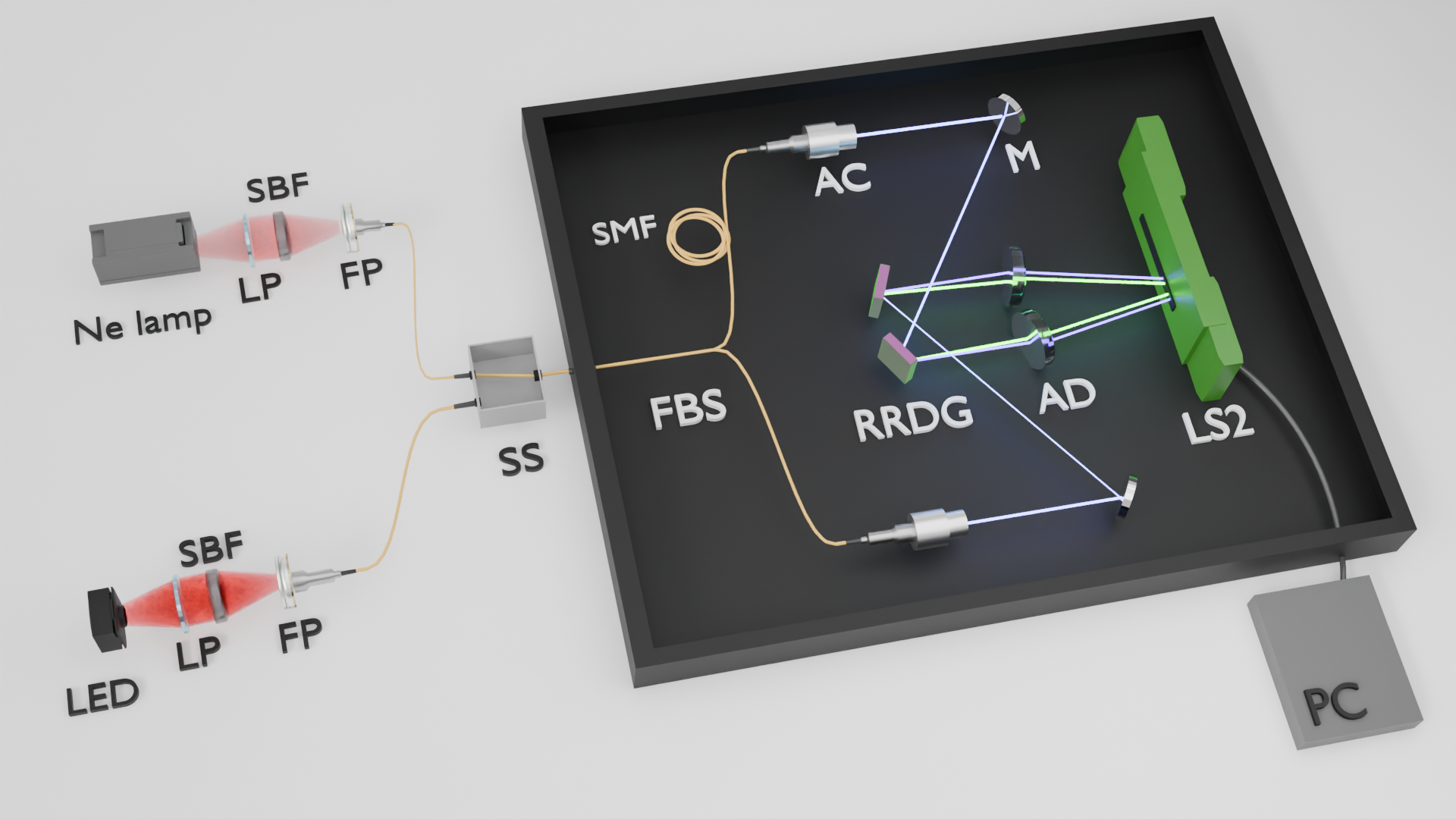}
\caption{Layout of dual spectrometer with two light sources: neon lamp and light-emitting diode (LED). For both sources, the light is polarized with a linear polarizer (LP) and filtered through a spectral bandpass filter (SBF) before entering a fiber port (FP), where it is coupled into a single-mode fiber. The source switch (SS) is essentially a fiber-to-fiber connecting sleeve, which allows for the selection of one of the two sources at a time. A 1-to-2 50:50 fiber-coupled beamsplitter (FBS) splits and sends the light into two arms of the spectrometer. One of the arms is delayed via an additional single-mode fiber (SMF). Both arms are connected to adjustable collimators (AC) that are used for focusing the beams of light. With the help of a mirror (M), the output beams are guided onto a ruled reflective diffraction grating (RRDG). The resulting spectrum is focused via an achromatic doublet (AD) onto the LinoSPAD2 sensor (LS2). Everything besides the two light sources and fiber-coupling components is located inside a light-tight enclosure. All fibers used in this setup are single-mode ones, including the beamsplitter.}
\label{fig:BlenderSpectrometer}
\end{figure}

We describe below a fast spectrometer with spectral and temporal resolutions of 40~pm and 40~ps, respectively, together with light sources used for the measurements and calibrations. Figure \ref{fig:BlenderSpectrometer} shows the layout of the spectrometer and two light sources used in the measurements. Similar spectrometers have already been described in our previous publications \cite{Nomerotski2023, Jirsa2024, Ferrantini2025}, so only a brief recap of their design and essential components follows. One of the main outcomes of this work is that we observed the photon bunching owing to the HBT effect for pairs of thermal photons from a broadband source that were matched in time and in frequency. The photons without frequency matching do not show this behavior.

\subsection{LinoSPAD2 detector}

The key part of our spectrometer is the LinoSPAD2 camera. In the LinoSPAD2 sensor \cite{bruschini2023, milanese2023linospad2}, each pixel is a single-photon avalanche photodiode (SPAD) with a $26.2 \times 26.2$ µm$^{2}$ size and 25.1\% native fill factor (can be improved with microlenses). The sensor consists of a linear $512\times1$ pixel array with a median dark count rate of $\sim 125$ cps/pixel, cross-talk of 0.2\% \cite{kulkov2024inter, kulkovspie2025} and final photon detection efficiency of about 30\% at 520~nm and 20\% at 640~nm for the sensor equipped with microlenses \cite{bruschini2023, Bruschini23_microlens} while operating at room temperature. After calibrations, the temporal resolution was determined to be equal to 40~ps (rms) employing an SPDC source of photon pairs \cite{SK_dissertation}. In this work, we employed 256 pixels (one half of the sensor).
LinoSPAD2 has already demonstrated good performance in detecting the HBT photon bunching signatures in our prior experiments with neon and argon spectral lines \cite{Jirsa2024, Ferrantini2025, kulkov2024inter}.

\subsection{Spectrometer}

The light entering the spectrometer is polarized and filtered via a $(640\pm5)$~nm bandpass filter before entering a fiber port. A 1-to-2 50:50 fiber-coupled beamsplitter splits the light into two beams, directing it to the two arms of the dual spectrometer. The outputs of the beamsplitter are connected to adjustable collimators that help with focusing and improving the spectral resolution. Each beam is then guided onto a ruled reflective diffraction grating with 1200 grooves/mm after reflection in a silver-coated mirror. The diffracted linear image of the spectrum within each arm of the dual spectrometer is focused through a 200 mm focal length lens onto the sensor. 
Every fiber used in this setup, including the fiber-coupled beamsplitter, is single-mode, and everything except the light sources is located inside a light-tight enclosure. The total system efficiency (including the detector) after entering the fiber port was estimated to be about 10\%.

To separate a possible cross-talk from the HBT effect, one of the two signal photons is delayed via an additional 1~m fiber in one of the dual spectrometer arms. This results in a shift of $\sim5$ ns of the HBT peak relative to $\Delta t = 0$ in the coincidence histograms \cite{kulkov2024inter, kulkovspie2025}.

\subsection{Light sources and spectral calibrations}

\begin{figure}
\centering
\includegraphics[width=1.0\textwidth]{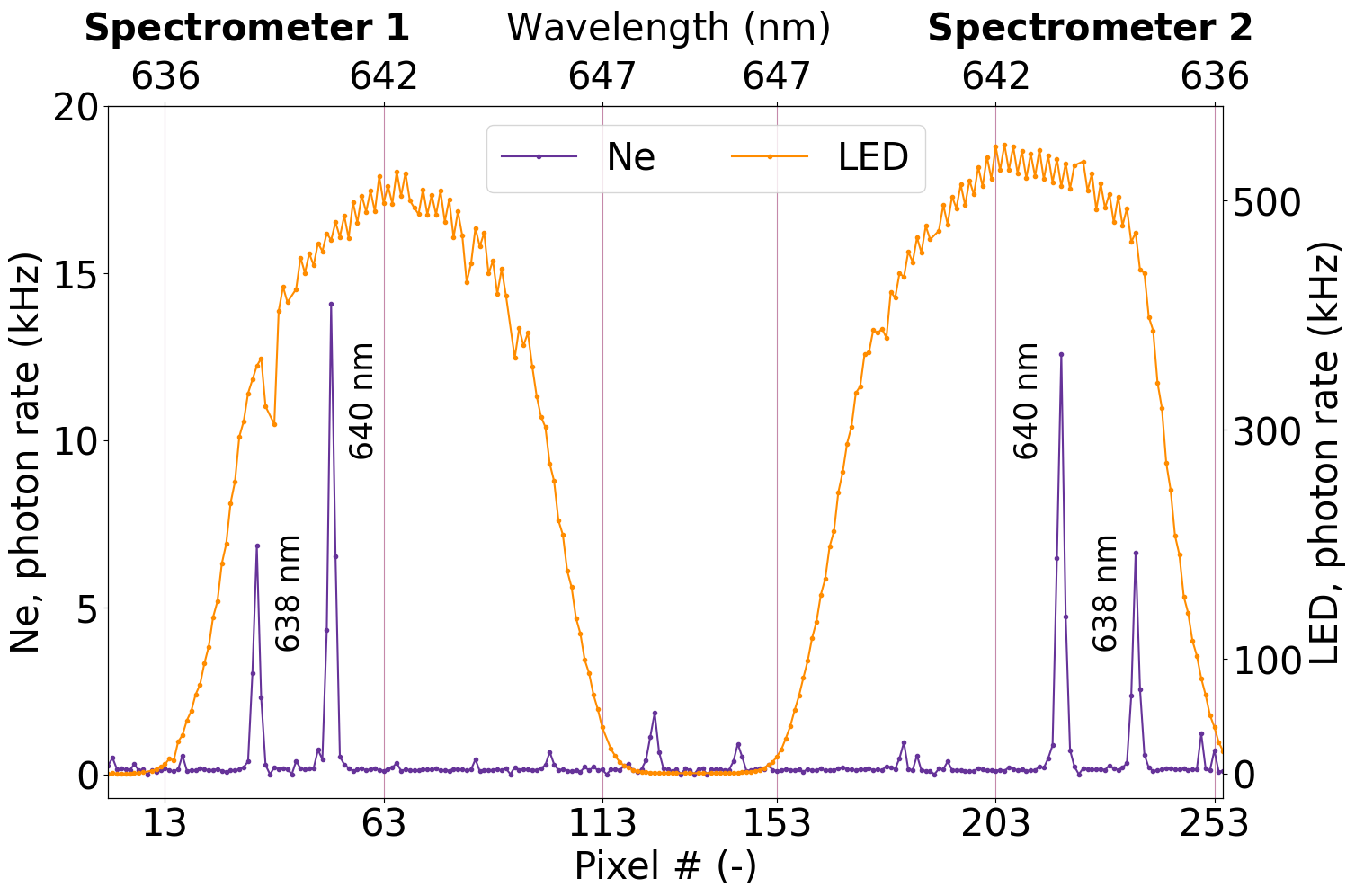}
\caption{Two 10 nm wide spectra in the dual spectrometer. The light source is an LED light passing through a $(640\pm5)$~nm bandpass filter. Two neon lines, 638.3 and 640.2~nm, measured in the same conditions as the broadband light, are overlaid with the broadband spectra, providing precise calibration of the spectrometers. Several weaker lines are also visible.}
\label{fig:twoSpectra}
\end{figure}

For the measurements presented here, we used two types of thermal light sources: broadband LED and neon lamp, both shown in Figure \ref{fig:BlenderSpectrometer}. Figure \ref{fig:twoSpectra} shows two mirrored instances of a 10 nm wide spectrum in the dual spectrometer. The spectrometer optics were aligned such that the two spectra appeared in the same sensor without overlap. This is directly exploiting the data-driven operation of the LinoSPAD2 single-photon detector, which can independently register multiple photons arriving at the same time with excellent temporal resolution.

The spectrometer was characterized using the neon emission spectrum~\cite{NIST_ASD}, which has a large number of narrow lines. Figure \ref{fig:twoSpectra} shows two neon spectra in the dual spectrometer. Two bright neon lines, 638.3 and 640.2 nm, are prominent in both spectrometers. Note that the wavelength is increasing in opposite directions for the two spectrometers due to the orientation of their optical elements. The two neon lines have well-defined wavelengths and can be used to calibrate the correspondence of the wavelength to position in the sensor. The resolution of the spectrometer as well as its scale was determined in a separate study \cite{Jirsa2024}, where we used the Ar spectrum that we projected onto the sensor of the LinoSPAD2 camera. Fitting with a Gaussian function all of the spectral lines that we measured, we determined the resolution of 0.04 nm (rms), while the scale was estimated using the distances between several known lines and was measured to be equal to 0.11 nm/pixel. Repeating the same procedure with Ne lines, in this work, we measured the same numbers --- as expected, since the setup is the same besides the change of the source.

\subsection{Data processing pipeline}

Acquisition of the dataset used for the analysis took nine hours, with a fraction of the detector live time with open shutter of about 10\%. The time-averaged (not accounting for downtime between acquisition cycles and for data transfer) maximum per-pixel photon count rate was
570 kHz (see Figure \ref{fig:twoSpectra}), while the total photon rate from all 256 pixels was approximately 71 MHz, corresponding to a data rate of $\sim270$ MB/s and resulting in 2 TB of raw data. 

In the data analysis, pairs of pixels from the two different spectrometers were selected, and their timestamps were compared, calculating delays between detections of pairs of photons. These timestamp differences are then used to build a histogram of coincidences, where a peak --- or enhancement in coincidences --- may correspond to either the cross-talk or HBT effects \cite{SK_dissertation}. The preprocessed timestamp differences were saved in independent datasets to separate this part of the analysis from its final phase involving the fitting. This resulted in another 2 TB of preprocessed data.

\section{Results}\label{sec2}

\subsection{HBT peaks}\label{subsec2}

\begin{figure}[h!]
\centering
\includegraphics[width=0.49\textwidth]{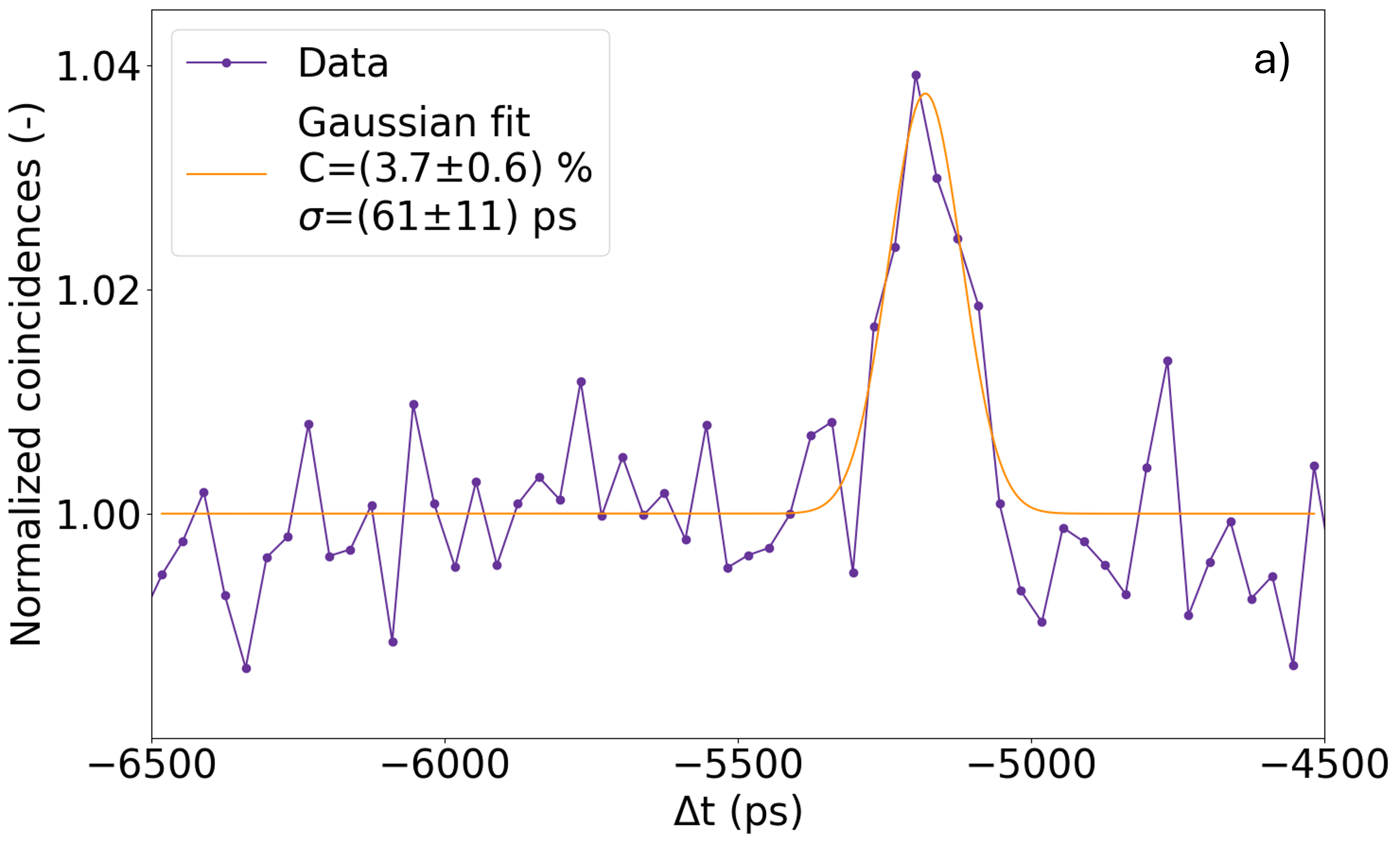}
\includegraphics[width=0.49\textwidth]{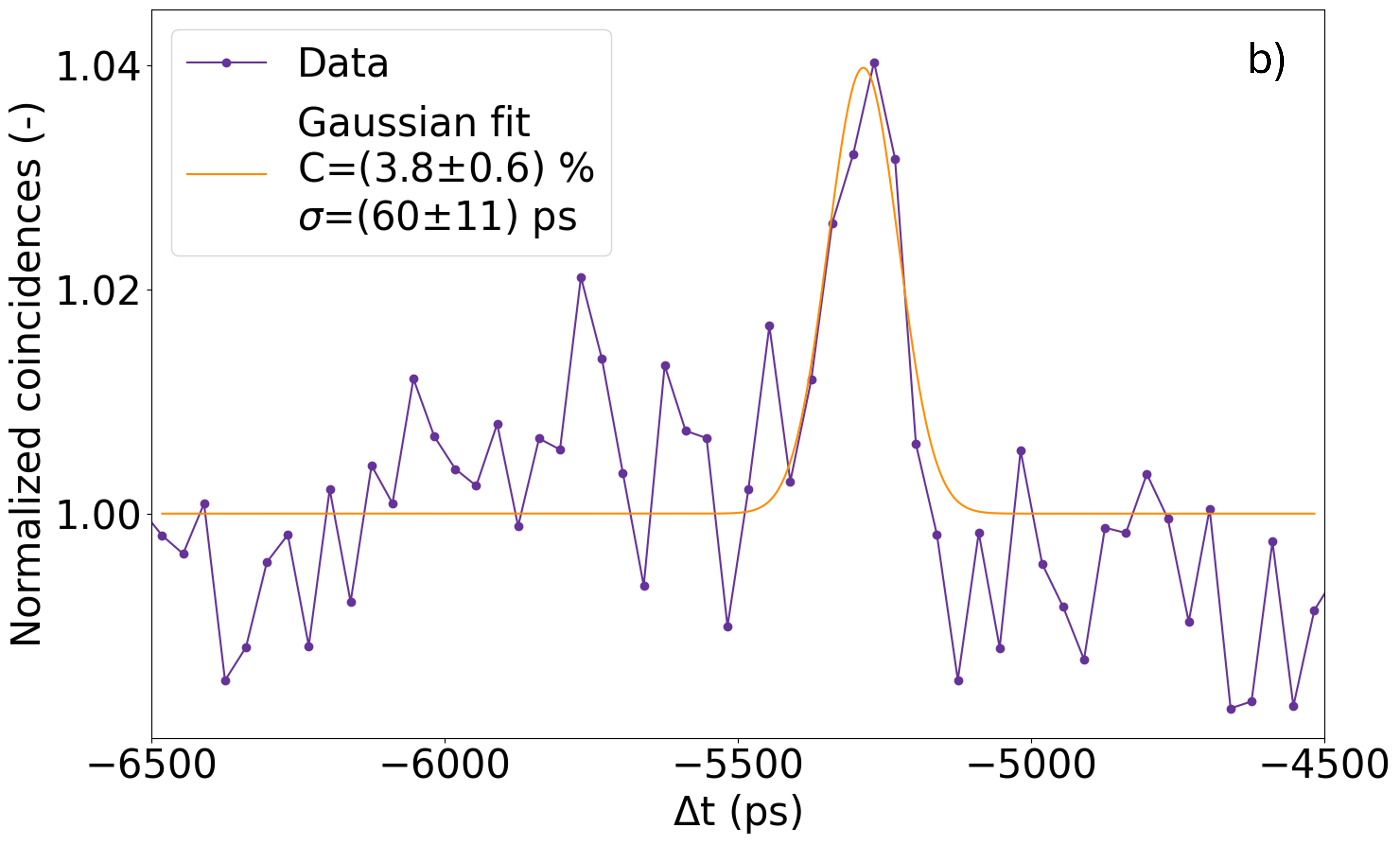}
\includegraphics[width=0.49\textwidth]{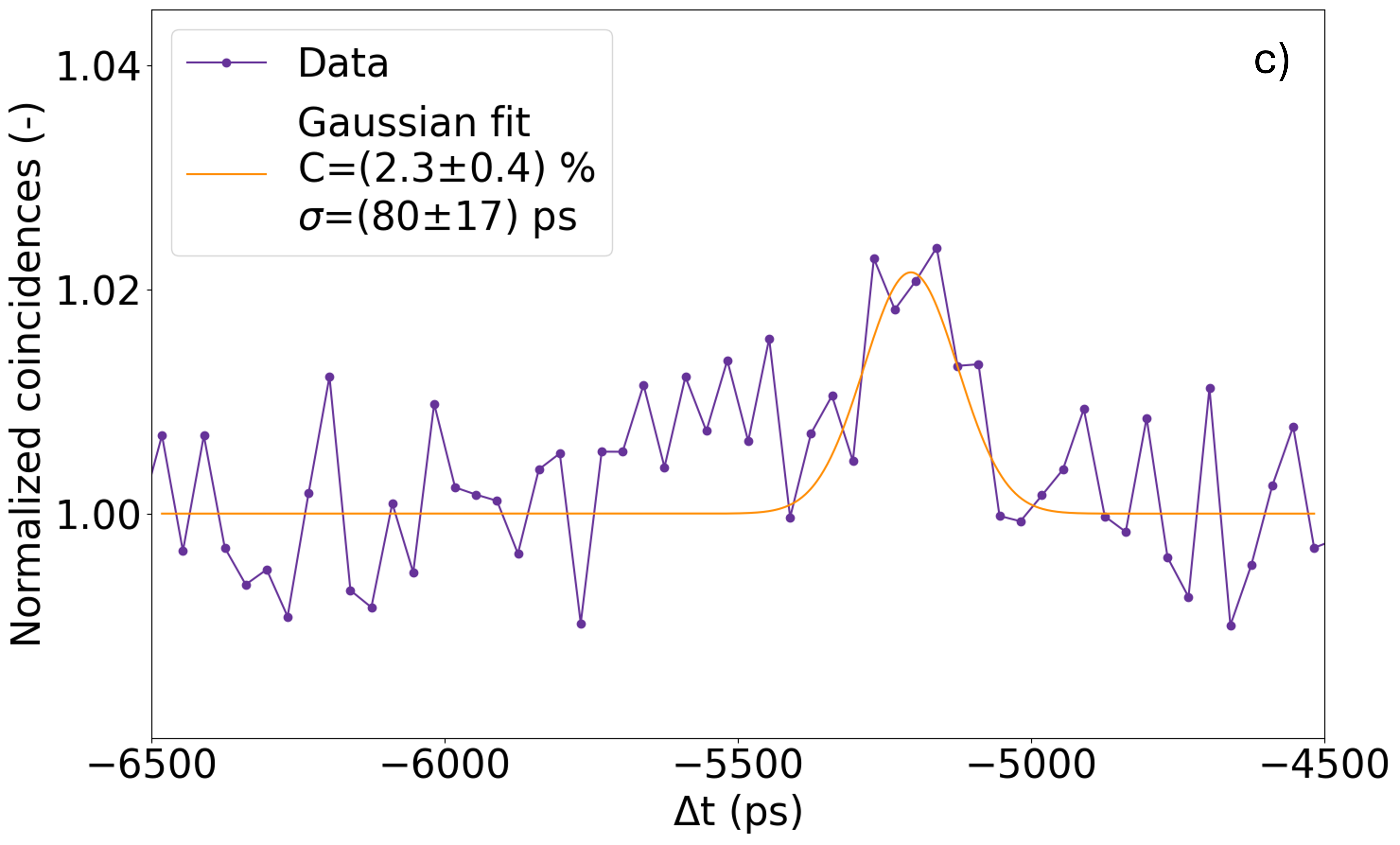}
\includegraphics[width=0.49\textwidth]{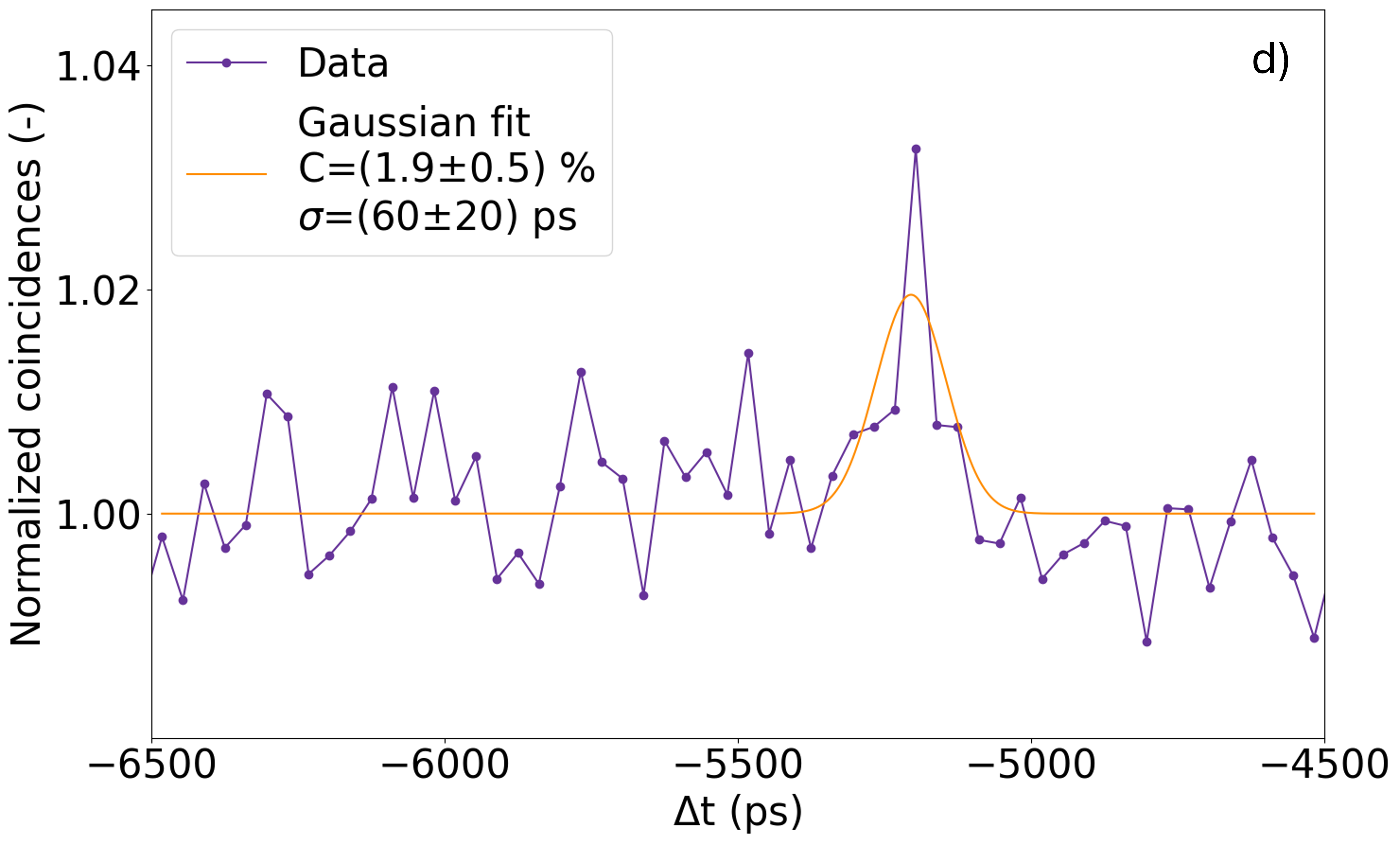}
\includegraphics[width=0.49\textwidth]{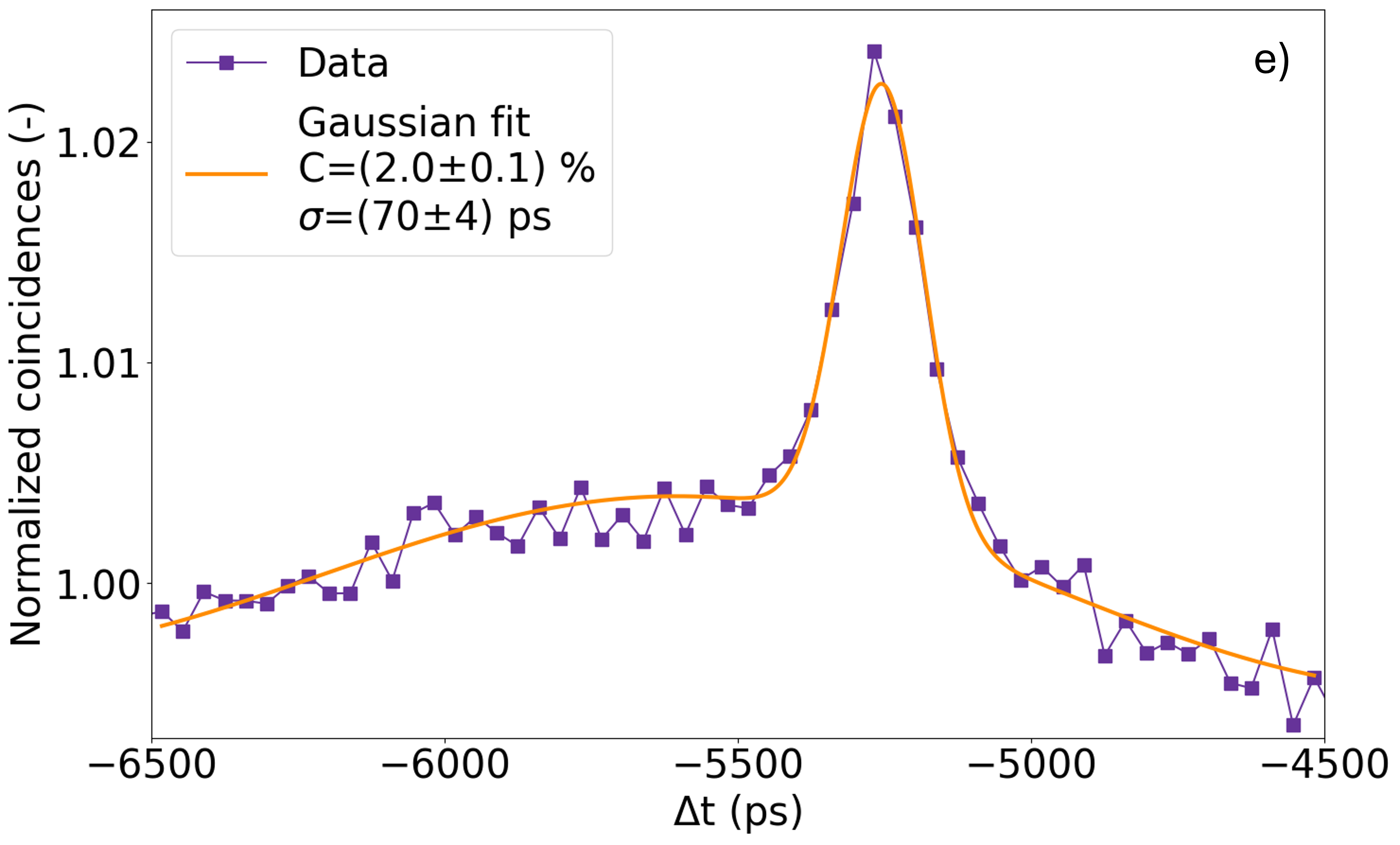}
\includegraphics[width=0.49\textwidth]{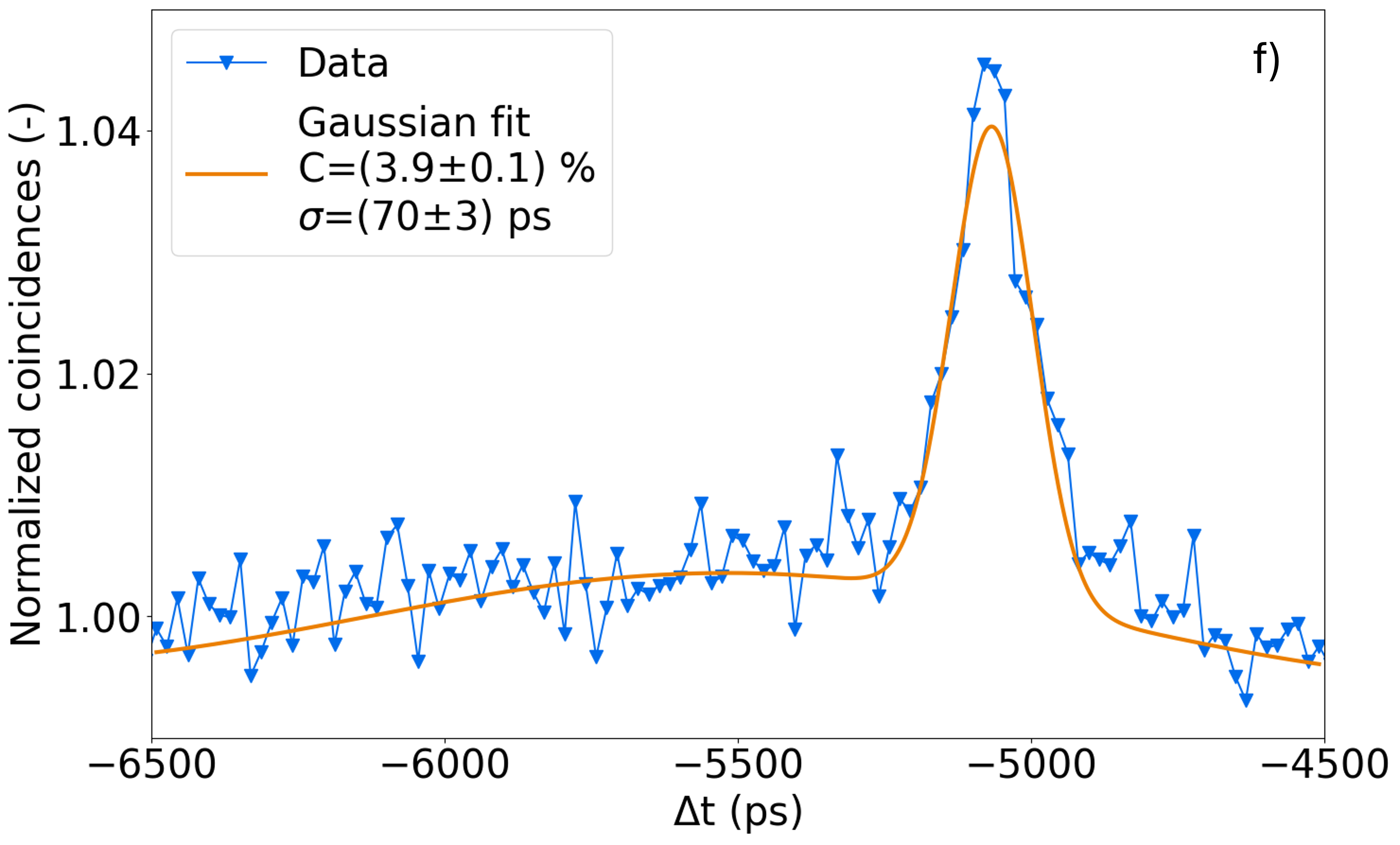}
\caption{a--d) Time difference distributions in two spectral bins in the dual spectrometer, four combinations in total (pixels 51 and 218, 90 and 179, 50 and 219, 46 and 223), with examples of different contrast values for HBT peaks determined by the fit. e) Averaged sum of all diagonal combinations where the two frequencies are expected to be equal. The peak contrast as determined by a fitting procedure is equal to $(2.0 \pm 0.1)$\% and sigma 70~ps; f) HBT peak after filtering with $(656.40\pm0.05)$ nm narrow band filter.}
\label{fig:HBTpeaksBins}
\end{figure}

In the data analysis, we select two pixels corresponding to two spectral bins from different spectrometers and find hits in those spectral bins with close timestamps. The HBT effect manifests itself as the enhancement of random time coincidences of thermal photons at a certain time difference, producing a so-called HBT peak. Theoretically, the maximum contrast of the peak, defined as the ratio of the peak height to the average background rate, can be equal to 100\%.
Figure \ref{fig:HBTpeaksBins} shows examples of HBT peaks after combining information from two pixels corresponding to two spectral bins (each from one of the spectrometer arms) with the same wavelength in our experiments.

To estimate the interference contrast in an unbiased manner, we developed a simple fitting procedure with a minimal number of free parameters and applied it uniformly to all combinations of pixel pairs. The HBT peak was fitted with a Gaussian function with its position predicted by the offset calibration \cite{kulkovspie2025} and constrained to $(-5200\pm200)$~ps of its nominal value. The Gauss sigma ($\sigma$) was constrained to the values from 60 to 80~ps, as its possible interval was determined in independent high-statistics measurements. The normalized background was set equal to one. The height of the peak --- or contrast ($C$) --- was constrained to the values from -10 to 10\%.

Typical cases presented in  Figure \ref{fig:HBTpeaksBins} have the peak contrast values from 1.9 to 3.8\%. 
Figure~\ref{fig:HBTpeaksBins}~e) shows the averaged sum of time difference distributions where all bin pairs of equal frequencies with expected non-zero signal are combined. The resulting distribution has a prominent HBT peak with $(2.0\pm0.1)$\% contrast and 70~ps sigma. In this case, non-ideal temporal offset calibrations would decrease the contrast and increase the width due to additional smearing, compared to the fits of individual pairs in Figure~\ref{fig:HBTpeaksBins}~a--d). We also note that in this case, we acquired enough statistics to observe that the background is not flat but has a slowly changing shape, which was accounted for by a sine term in the fit for this particular case.
Figure \ref{fig:HBTpeaksBins}~f) also shows the HBT peak for the LED broadband light after an ultra-narrowband $(656.40\pm0.05)$ nm filter was used without a spectrometer in a different set of measurements. Note that the width of the filter approximately matches the bin width in the spectrometer, resulting in similar contrasts. The measured contrast, in this case, is $(3.9\pm0.1)$\%.

\subsection{Matrix of contrasts}\label{subsec2}

\begin{figure}[h!]
\centering
\includegraphics[width=1\textwidth]{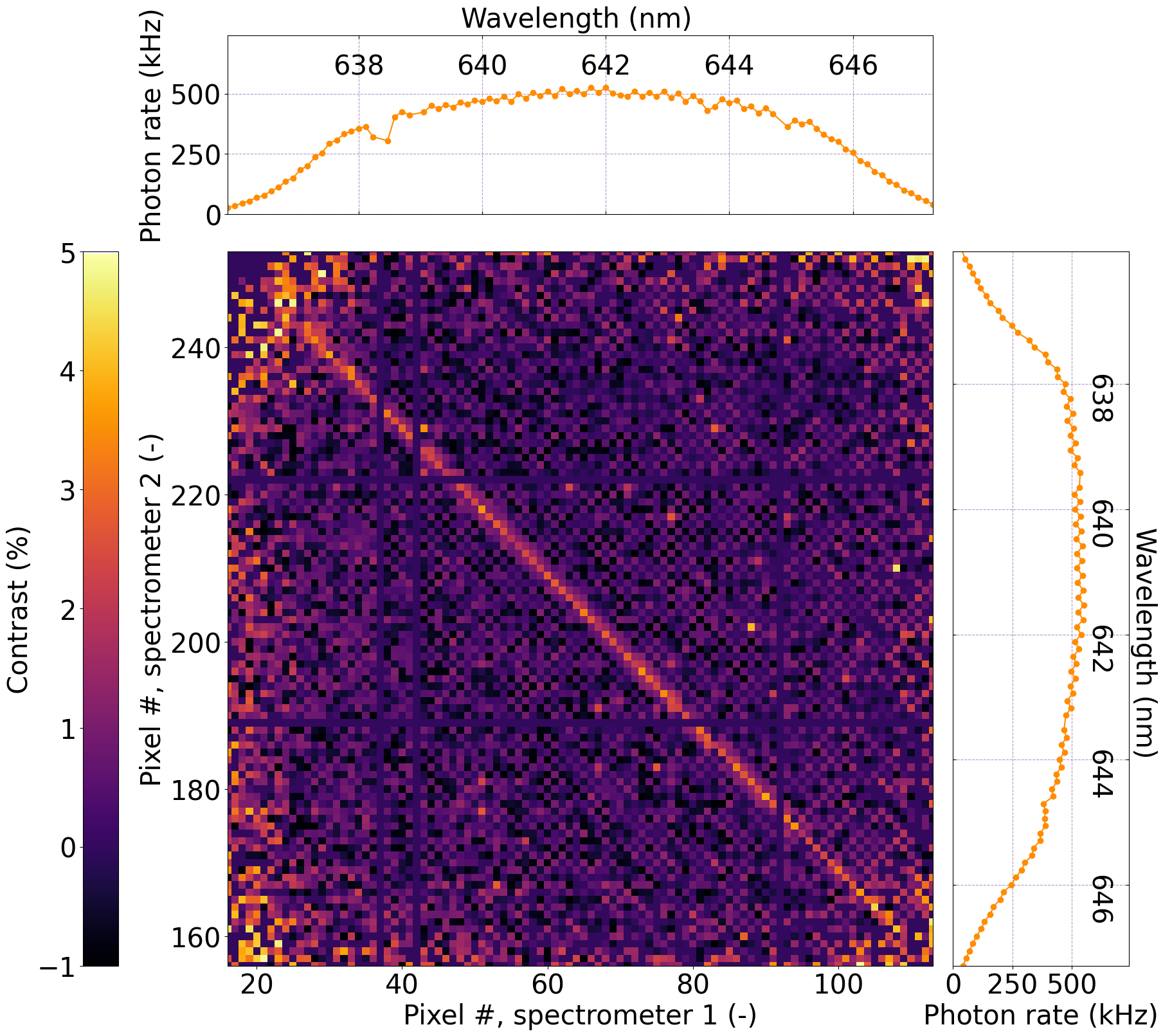}
\caption{Summary 2D histogram of contrasts extracted by the fitter from histograms of coincidences for each pair of spectral bins. The 1D plots above and on the right show the two spectra in the dual spectrometer. The diagonal corresponding to the spectral bins matched in frequency clearly shows the non-zero contrasts of the observed HBT peaks. As the edges of the two spectra have very low photon rates, this translates to large statistical fluctuations near the 2D histogram edges. The color scale for contrast values was limited from $-1$\% to 5\% to remove a small number of outliers on the edges. Empty vertical and horizontal lines correspond to several noisy, and therefore masked pixels, and also to dead pixels.}
\label{fig:matrix}
\end{figure}

Finally, Figure \ref{fig:matrix} shows the full matrix of contrasts for all combinations of pairs of spectral bins in a matrix of $100 \times 100$ pixels, so 10,000 combinations in total. The same fitting procedure that was used for Fig. \ref{fig:HBTpeaksBins}~a)--d) was applied to all pairs of pixels shown here with the same parameter estimation and limitations. We expect HBT peaks to appear only for the cases of equal frequencies for combinations of spectral bins that belong to the diagonal of this matrix, as is indeed apparent here. The contrasts were estimated by fitting the time difference distribution as described in detail in the previous subsection. In the case of the off-diagonal elements, where the HBT peaks were not expected, the contrast was close to zero, while for the diagonal elements with HBT peaks, the contrast is about 2 to 4\%. We consider this result a remarkable confirmation of expected behavior for the HBT effect massively multiplexed in frequency, and also a sound quality proof of our instrument.

\begin{figure}[h!]
    \centering
    \includegraphics[width=0.8\linewidth]{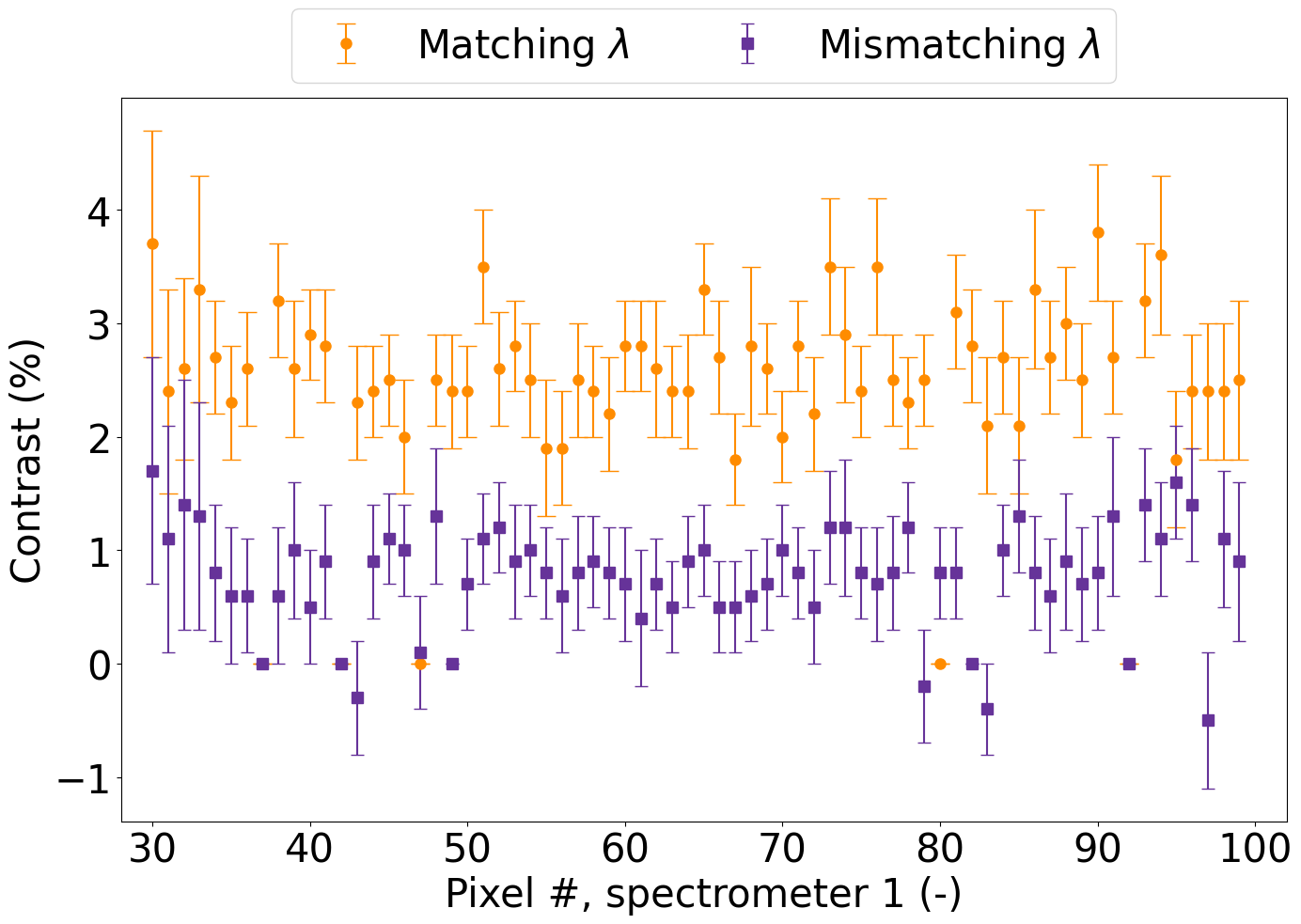}
    \caption{Contrast values along two selected diagonals: the HBT diagonal corresponding to equal wavelengths and a diagonal offset by 2 spectral bins (mismatched wavelengths).}
    \label{fig:contrast_error}
\end{figure}

Figure \ref{fig:contrast_error} shows contrasts for two different diagonals: the main one corresponding to the matching spectral bins showing contrasts of the HBT peaks and one shifted 2 spectral bins away, corresponding to mismatching bins. A mismatch of just $\sim0.2$ nm in the wavelength of the two photons leads to a loss of the photon bunching.

\section{Discussion}\label{sec12}

In the following section, we discuss practical implications of our results for astrometric resolution and quantum entanglement-swapping protocols, highlighting pathways for further improvements through advanced detector technologies.

\textbf{Astrometric resolution}: The presented results are an important step forward for the stellar intensity interferometry and quantum-assisted schemes of optical interferometers \cite{Crawford2023, Brown2023} because the spectrometer can be directly employed for spectral binning in the corresponding experiments, with each bin effectively performing an independent measurement, which can be combined with other measurements to reduce the uncertainty. In the case of the tested spectrometer, we can combine about 200 pairs of spectral bins where the contrast is non-zero, corresponding to three adjacent diagonal lines. In the considered two-photon interferometer, the opening angle between two stars $\Delta\theta$ is extracted from the population of HBT peaks measured for different combinations of detectors in two stations of the interferometer \cite{Stankus2022}. The associated uncertainty of the opening angle can be expressed as in \cite{Stankus2022}, showing dependence on the most important variables:

\begin{equation}
    \sigma \left[ \Delta \theta \right] \propto
\frac{\lambda}{B} \, 
\frac{1}{V} \,
\frac{1}{T}
\frac{1}{\sqrt{\bar{n} T}}.
\label{eq:stdv_deltatheta}
\end{equation}

It is proportional to wavelength $\lambda$ and inversely proportional to baseline (i.e., distance) $B$ between interferometer stations, allowing us to gain from longer baselines. The temporal and spectral resolutions of the spectrometer are accounted for by the visibility $V$, while other photon collection and detection parameters, such as the size of the mirrors and the detection efficiency, are accounted for in the detected rate of photons $\bar{n}$.  With $T$ being the length of observation period we also note that the overall $T^{-3/2}$ dependence is more favorable than simply the expected $T^{-1/2}$ gain from photon pair statistics, reflecting the advantage of being able to use the measurement of a fringe oscillation rate due to the Earth rotation as proposed in \cite{Stankus2022} and elaborated in \cite{Crawford2023}. 

The contrasts presented in Figure \ref{fig:matrix} have the same definition as $V$. In the presented study, we directly measured the visibility for each pair of spectral bins and, therefore, can determine the improved uncertainty $\sigma \left[ \Delta \theta \right]$ by combining visibility measurements in multiple bins, assuming that all other parameters (baselines, photon rates, observing time) in Eq.~\ref{eq:stdv_deltatheta} are the same for different spectrometers in both observing stations. Employing 70 spectral bins from the main diagonal would allow us to decrease the astrometric uncertainty by a factor of 8.3, while using extra bins from two adjacent diagonals, which are expected to have non-zero visibility because of the partial spectral overlap of the neighboring spectral bins, improved this value to 10.0. We consider this an important step to demonstrate the advantages of frequency multiplexing and the viability of the approach for this application. We also note that the number of spectral bins can be scaled up by a large factor employing larger and multiple sensors, and that it is not unusual to have tens of thousands of spectral channels on modern astronomical spectrographs \cite{martini2018overview}.

\textbf{Heisenberg-Gabor limit}: It is instructive to compare the achieved resolution to the Heisenberg Uncertainty Principle limit or, equivalently, to the Heisenberg-Gabor limit. We note that the limits do not constrain the resolution for one of the observables alone, but restrict the product of the uncertainty of two observables, taking the form of:
\begin{equation}
    \Delta E \Delta t \geq \frac{\hbar}{2} ;~~~~ \Delta f \Delta t \geq \frac{1}{4\pi},
\end{equation}
where $\hbar$ is Planck's reduced constant, $\Delta E$, $\Delta f$, and $\Delta t$ are respectively the standard deviations for measurements of energy, frequency, and time. 
In our case of 640~nm wavelength and assuming 40~ps \& 40~pm rms resolutions \cite{Jirsa2024}, $\Delta f \Delta t$ is a factor of 14.7 larger than the Heisenberg-Gabor limit, interpreting the spectral resolution uncertainty as frequency uncertainty. In general, this comparison is a useful benchmark for the indistinguishability of photons if they hit the same spectral bins in the two spectrometers. Experimentally, this means that the contrast of the corresponding HBT peak cannot be larger than $100\%/14.7=6.8\%$, see a complete mathematical treatment of the contrast derivation in Appendix A.
The measured contrast of about 3\% is indeed limited by the finite spectral and temporal resolutions of the spectrometer, and it can be further diluted by the relative calibrations of the spectrometers and other imperfections.

\textbf{Entanglement swapping}: Frequency multiplexing offers a powerful route to scale up entanglement swapping protocols in quantum communication, where SPDC sources are widely used to generate entangled photon pairs. A key limitation of SPDC is the generation of multi-pair events, which degrade quantum correlations; for QKD, the double-pair emission probability must typically remain below 5\%, with even tighter constraints for other quantum applications  \cite{Yuri_1, Yuri_2}.

To enhance the rate of successful entanglement swaps, one promising strategy is to employ a broadband entanglement source, which can be viewed as comprising many narrowband emitters, multiplexed across frequency. The spectrometer described here can function as a multichannel entanglement-swapping platform, simultaneously receiving inputs from two entangled-photon sources. Each half of the sensor acts as an independent spectrometer, with pixels symmetrically paired across the array to monitor the same wavelength. Our configuration effectively enables up to 128 parallel Bell-state measurements (assuming 256 pixels in total are available) --- corresponding to the number of spectral channels --- thus increasing the entanglement swap rate by up to two orders of magnitude.

Given excellent timing resolution of the spectrometer, this architecture is particularly compatible with phase-encoded time-bin qubits \cite{Zbinden_Gisin_timebin}, in which each qubit is encoded as a photon in a coherent superposition of early and late time bins. Entanglement swapping is then identified by two-photon coincidence events within the same spectral bin \cite{Zbinden_Gisin_swap}, which inform the remote parties of the spectral channel hosting the successfully entangled pair for subsequent quantum operations.The main limitation of the presented spectrometer for this application is its low overall system efficiency, estimated at approximately 
10\% and dominated by the SPAD photon detection efficiency. For two-photon processes, this limitation is particularly severe because the relevant coincidence rates scale with the square of efficiency. Nevertheless, these results represent an important step toward spectral multiplexing for entanglement swapping, demonstrating the feasibility of the approach. As detector technologies continue to improve, higher-efficiency implementations are expected to make this method a competitive and practical solution.

\textbf{Current state of the art and future advancements:} In Table \ref{tab1:comparison} we compare our results with those of other published fast spectrometers, focusing on the operational wavelength $\lambda$, spectral and temporal resolutions $\Delta \lambda$ \& $\Delta t$, multiplication factor with respect to the Heisenberg-Gabor limit (i.e. how far the product of resolutions is above the limit), number of spectral channels, and the possibility to measure multiple photons in parallel. All examples in the table are far inferior to our results in terms of the product of two resolutions and, in some cases, in the number of spectral channels.

\begin{table}[htp]
\centering
\footnotesize
\setlength{\tabcolsep}{4pt}
\renewcommand{\arraystretch}{1.05}

\caption{Comparison of fast single-photon spectrometers. Column "HGL" corresponds to the multiplication factor for the Heisenberg-Gabor limit, i.e. how far the product of resolutions is above the limit.}
\label{tab1:comparison}

\begin{tabular}{ccccccc}
\toprule
$\lambda$ (nm) &
$\Delta \lambda$ (nm) &
$\Delta t$ (ns) &
HGL 
&
\# of chan. &
multi-photon &
ref. \\
\midrule

640  & 0.04  & 0.04  & 14.7     & 256  &  yes       & this work \\
531 & 1.7 & 0.02 & 486     & 8   & yes    & \cite{Dilena2025} \\
780 & 0.33 & 0.38 & 781     & 5   & yes        & \cite{tolila2024increasing} \\
600 & 7.0 & 0.03 & 2420     & 200   & no        & \cite{cheng2019broadband} \\
500  & 2.0 & 0.11   & 3300   & 32     & yes         & \cite{johnsen2014time} \\
800  & 0.5   & 7.30  & 22300  & 230   &yes         & \cite{farella2024spectral} \\
800  & 0.7     & 6.00   & 24700    & 256  & yes     & \cite{zhang2021high} \\
\bottomrule
\end{tabular}
\end{table}

An important case for comparison is presented in \cite{cheng2019broadband}, which employed a different approach. It used a single channel of the long superconducting nanowire single-photon detector (SNSPD) with a diffraction grating to disperse light along the nanowire, determining the wavelength from the time difference of signal arrival at its two ends and leveraging the detector’s excellent timing resolution. However, the spectral resolution achieved was much poorer than in our work, with long and irregular tails in the resolution function.

Future advancements in fast sensor technologies are expected to significantly enhance temporal resolution, potentially achieving resolutions as fine as 5 ps rms for SPAD devices, as recently demonstrated in  \cite{Gramuglia2022}. Hybrid solutions with a SPAD sensor attached to a time-stamping chip are also being investigated as a route to large two-dimensional fast imagers \cite{Llopart2022,Hogenbirk2026, picopix,timespot}. Similarly, SNSPDs are rapidly advancing toward temporal resolutions in the single-digit picosecond regime. A notable advantage of this type of detector is its superior quantum efficiency compared to SPADs, which is important in many applications. However, these detectors currently require cryogenic temperatures, and their scalability to multi-channel and two-dimensional detector arrays has not yet been fully established \cite{Cheng2019, Korzh2020, oripov2023camera}.

Spectral resolution improvements down to a few picometers can be realized by employing an Echelle spectrometer configuration \cite{Nagaoka1923}. In this arrangement, high-order diffraction provides enhanced spectral resolution, while a first-order grating vertically separates overlapping spectral orders. The resulting output, a sequence of parallel spectral stripes, necessitates the use of a two-dimensional array of single-photon detectors to achieve high spectral resolution across an extensive wavelength range.

Collectively, these enhancements in temporal and spectral resolutions could restore visibility close to the theoretical limit, 100\%, increasing it with respect to the present level by a factor of 20--40. This will translate to the same scale improvement of achievable astrometric resolution for the quantum-assisted astrometry applications and improved performance of the Bell-state measurements in quantum networking applications.

Ultimately, our realization of massively frequency-multiplexed two-photon correlation measurements presents a transformative advancement, with the potential to accelerate the development of quantum sensing and quantum communication technologies by orders of magnitude.

\section{Conclusion}\label{sec13}

We have demonstrated the first observation of the HBT effect for broadband light dispersed in a fast spectrometer across a wide, continuous spectral range of 10 nm, utilizing a large number, of the order of 100, of spectral channels. After precise calibration, our spectrometer achieved exceptional temporal and spectral resolutions of 40 ps and 40 pm, respectively. This enabled the observation of thermal photon bunching with an average contrast of approximately 3\%, consistent with theoretical expectations. The achieved temporal and spectral filtering approaches the Heisenberg-Gabor limit, making our spectrometer highly applicable to diverse fields such as quantum communications, quantum metrology, spectroscopy, and quantum-assisted telescopy.

Analyzing the spectrometer’s performance specifically in the context of phase-sensitive intensity interferometry, we found that combining about 200 spectral bin combinations exhibiting non-zero contrast could enhance astrometric precision by a factor of 10. Furthermore, we have outlined the significant potential gains achievable through entanglement swapping in quantum networks, where frequency multiplexing could dramatically enhance the throughput and robustness of entangled photon distribution.

Future advancements in temporal and spectral resolution promise further significant improvements, potentially enabling near-ideal visibility of two-photon interference phenomena. Ultimately, our realization of massively frequency-multiplexed two-photon correlation measurements presents a transformative advancement, with the potential to accelerate the development of quantum sensing and quantum communication technologies by orders of magnitude.

\appendix

\section{Appendix: Spectrally and temporally resolved thermal two-photon correlations}

We model the broadband source as a stationary multimode thermal state with
\begin{equation}
\langle \hat a^\dagger(\omega)\hat a(\omega')\rangle
=
\overline{n}\,S(\omega)\,\delta(\omega-\omega'),
\qquad
\int_{-\infty}^{\infty} S(\omega)\,d\omega = 1,
\end{equation}
where $\overline{n}$ is the total count rate and $S(\omega)$ is the normalized source spectrum, $\hat{a}^{\dagger}(\omega)$ and $\hat{a}$ are standard quantum optical creation and annihilation operators of a photon with frequency $\omega$.

To distinguish the coarse optical filtering performed before the spectrometer from the intrinsic spectral resolution of the detector array, we introduce two different transfer functions:
\begin{itemize}
    \item $B(\omega)$ --- amplitude transfer function of the external optical filter placed before the detector;
    \item $H_j(\omega)$ --- amplitude spectral response of the $j$-th spectrometer channel (or detector bin).
\end{itemize}
The effective detected field operator in channel $j$ is then
\begin{equation}
\hat A_j(t)
=
\int_{-\infty}^{\infty} d\omega\,K_j(\omega)\hat a(\omega)e^{-i\omega t},
\qquad
K_j(\omega)=B(\omega)H_j(\omega).
\label{eq:Aj_def}
\end{equation}
The mean count rate in channel $j$ is
\begin{equation}
N_j
=
\langle \hat A_j^\dagger(t)\hat A_j(t)\rangle
=
\overline{n}\int_{-\infty}^{\infty} S(\omega)|K_j(\omega)|^2\,d\omega.
\label{eq:Nj_def}
\end{equation}
This equation allows one to estimate how many photons are lost due to filtration. For two channels $j$ and $k$, the first-order correlation function is
\begin{equation}
G^{(1)}_{jk}(\tau)
=
\langle \hat A_j^\dagger(t)\hat A_k(t+\tau)\rangle
=
\overline{n}\int_{-\infty}^{\infty}
S(\omega)\,K_j^*(\omega)K_k(\omega)e^{-i\omega\tau}\,d\omega,
\label{eq:G1jk_def}
\end{equation}
where $\tau=t_2-t_1$.

Canonically, thermal sources are usually described via a multimode density operator with a Gaussian-like Glauber-Sudarshan $P$-function, where the fourth-order moment factorizes according to the standard Glauber rule:
\begin{equation}
G^{(2)}_{jk}(\tau)
=
\langle
\hat A_j^\dagger(t)\hat A_k^\dagger(t+\tau)
\hat A_k(t+\tau)\hat A_j(t)
\rangle
=
N_jN_k+\bigl|G^{(1)}_{jk}(\tau)\bigr|^2.
\label{eq:G2jk_factorized}
\end{equation}
Therefore, the normalized second-order correlation function is
\begin{equation}
g^{(2)}_{jk}(\tau)
=
\frac{G^{(2)}_{jk}(\tau)}{N_jN_k}
=
1+\frac{\bigl|G^{(1)}_{jk}(\tau)\bigr|^2}{N_jN_k}.
\label{eq:g2jk_general}
\end{equation}

Equation~(\ref{eq:g2jk_general}) immediately explains the main experimental observation:
for matched spectral bins ($j=k$), the overlap in Eq.~(\ref{eq:G1jk_def}) is maximal and the HBT bunching peak is observed, whereas for mismatched bins ($j\neq k$), the spectral overlap becomes small and the bunching term is suppressed.

\paragraph{Matched spectral bins.}
Below we explore further the diagonal case when $j=k$. Then Eqs.~(\ref{eq:G1jk_def})--(\ref{eq:g2jk_general}) reduce to
\begin{equation}
g^{(2)}_{jj}(\tau)
=
1+
\left|
\frac{
\int_{-\infty}^{\infty} S(\omega)|K_j(\omega)|^2 e^{-i\omega\tau}\,d\omega
}{
\int_{-\infty}^{\infty} S(\omega)|K_j(\omega)|^2\,d\omega
}
\right|^2 .
\label{eq:g2jj_general}
\end{equation}

Assume now that the source spectrum, the external optical filter, and the detector channel response are Gaussian:
\begin{equation}
S(\omega)\propto
\exp\!\left[-\frac{(\omega-\omega_s)^2}{2\Delta\omega_{\mathrm{src}}^2}\right],
\end{equation}
\begin{equation}
|B(\omega)|^2\propto
\exp\!\left[-\frac{(\omega-\omega_B)^2}{2\delta\omega_{\mathrm{ext}}^2}\right],
\end{equation}
\begin{equation}
|H_j(\omega)|^2\propto
\exp\!\left[-\frac{(\omega-\omega_j)^2}{2\delta\omega_{\mathrm{det},j}^2}\right].
\end{equation}
Then the product
\begin{equation}
S(\omega)|B(\omega)|^2|H_j(\omega)|^2
\end{equation}
is again Gaussian, with effective rms width
\begin{equation}
\Sigma_j^2
=
\left(
\frac{1}{\Delta\omega_{\mathrm{src}}^2}
+
\frac{1}{\delta\omega_{\mathrm{ext}}^2}
+
\frac{1}{\delta\omega_{\mathrm{det},j}^2}
\right)^{-1}.
\label{eq:Sigmaj_def}
\end{equation}
Hence Eq.~(\ref{eq:g2jj_general}) becomes
\begin{equation}
g^{(2)}_{jj,\mathrm{id}}(\tau)
=
1+\exp\!\left(-\Sigma_j^2\tau^2\right),
\label{eq:g2jj_ideal}
\end{equation}
where $\rm{id}$ indicates an ideal measurement with ideal temporal resolution.

This expression reproduces the expected limits. In particular, if the external optical filter is broad compared to the detector spectral channel width, i.e,
\begin{equation}
\delta\omega_{\mathrm{ext}}\gg \delta\omega_{\mathrm{det},j}.
\end{equation}
Then the detector channel dominates the spectral resolution and
\begin{equation}
\Sigma_j^2
\approx
\left(
\frac{1}{\Delta\omega_{\mathrm{src}}^2}
+
\frac{1}{\delta\omega_{\mathrm{det},j}^2}
\right)^{-1}.
\end{equation}

\paragraph{Finite temporal resolution.}
The temporal resolution of the detector array can be modeled by convolving the ideal bunching term with the instrumental response in the delay variable. Let
\begin{equation}
R(\tau)
=
\frac{1}{\sqrt{2\pi}\sigma_\tau}
\exp\!\left(-\frac{\tau^2}{2\sigma_\tau^2}\right),
\qquad
\int_{-\infty}^{\infty}R(\tau)\,d\tau = 1,
\end{equation}
where $\sigma_\tau$ is the rms resolution of the measured time-difference histogram. The measured normalized second-order correlation is then
\begin{equation}
g^{(2)}_{jj,\mathrm{M}}(\tau)-1
=
\int_{-\infty}^{\infty}
R(\tau-\tau')
\left[g^{(2)}_{jj,\mathrm{id}}(\tau')-1\right]\,d\tau'.
\label{eq:g2_conv}
\end{equation}
Substituting Eq.~(\ref{eq:g2jj_ideal}) into Eq.~(\ref{eq:g2_conv}) and evaluating the Gaussian convolution gives
\begin{equation}
g^{(2)}_{jj,\mathrm{M}}(\tau)
=
1+\mathcal{C}_j\exp\!\left(-\Sigma_{j,\mathrm{M}}^2\tau^2\right),
\label{eq:g2_measured_final}
\end{equation}
with contrast
\begin{equation}
\mathcal{C}_j
=
\frac{1}{\sqrt{1+2\Sigma_j^2\sigma_\tau^2}},
\qquad
\Sigma_{j,\mathrm{M}}
=
\frac{\Sigma_j}{\sqrt{1+2\Sigma_j^2\sigma_\tau^2}}.
\label{eq:contrast_final}
\end{equation}
If the two spectrometer arms have independent rms timing jitters $\sigma_{t,1}$ and $\sigma_{t,2}$, then
\begin{equation}
\sigma_\tau^2=\sigma_{t,1}^2+\sigma_{t,2}^2.
\end{equation}
For identical timing resolution in both arms, $\sigma_{t,1}=\sigma_{t,2}=\sigma_t$, one obtains
\begin{equation}
\sigma_\tau=\sqrt{2}\,\sigma_t,
\end{equation}
and therefore
\begin{equation}
\mathcal{C}_j
=
\frac{1}{\sqrt{1+4\Sigma_j^2\sigma_t^2}},
\qquad
\Sigma_{j,\mathrm{M}}
=
\frac{\Sigma_j}{\sqrt{1+4\Sigma_j^2\sigma_t^2}}.
\label{eq:contrast_identical_arms}
\end{equation}

\paragraph{Relation to the Heisenberg--Gabor limit.}
In the experimentally relevant regime where the source is broader than a single detector spectral channel and the external optical filter is also broad on that scale, one has
\begin{equation}
\Sigma_j \approx \delta\omega_{\mathrm{det},j} = 2\pi\Delta f,
\qquad
\mathcal{C}_j \approx \frac{1}{\sqrt{1+\left(4\pi \Delta f\,\Delta t\right)^2}},
\end{equation}
with asymptotic value for $4\pi\Delta f\,\Delta t\gg1$:
\begin{equation}
\mathcal{C}_j
\approx
\frac{1}{4\pi\Delta f\,\Delta t}.
\label{eq:C_asymptotic}
\end{equation}
Thus, the factor by which the time--frequency resolution product exceeds the Heisenberg--Gabor limit provides a direct upper estimate for the observable HBT contrast in the Gaussian model. For the present experimental setup with $\Delta t=40~\mathrm{ps}$, $\Delta\lambda=40~\mathrm{pm}$ and $\lambda = 640$~nm,  this factor is equal to 14.7,  yielding:
\begin{equation}
\mathcal{C}_j \approx \frac{100\%}{14.7}\approx 6.8\%.
\end{equation}

\subsection*{Disclosures}

Edoardo Charbon is a co-founder of NovoViz. NovoViz has not been involved in this work or in the drafting of this document.
The authors declare that they have no conflict of interest. 

\subsection*{Code and Data Availability}

The raw measured data set is over 2 TB in size, and the preprocessed data that is used for histogram building is also approximately 2 TB in size, making it difficult to share online. Nonetheless, the data is archived and can be provided upon reasonable request. The analysis codes used are part of the open-source library daplis (available at https://github.com/rngKomorebi/daplis), developed by S. Kulkov.

\acknowledgments

This research was supported by the Czech Science Foundation (GACR) under Project No. 25-15534M and the Grant Agency of the Czech Technical University in Prague, Grant No. SGS24/\allowbreak 063/\allowbreak OHK4/\allowbreak 1T/\allowbreak 14. This work was also supported by the EPFL internal project “High-speed multimodal super-resolution microscopy with SPAD arrays” and DOE/LLNL project “The 3DQ Microscope”. 
The authors thank the Prague Observatory and Jakub Rozehnal for providing some of the optical equipment used in the experiments.

\bibliographystyle{JHEP}
\bibliography{biblio}

\end{document}